\documentclass[reqno,12pt]{iopart}
\usepackage{iopams}
\newcommand{\be}{\begin{equation}}
\newcommand{\ee}{\end{equation}}
\newcommand{\bea}{\begin{eqnarray}}
\newcommand{\eea}{\end{eqnarray}}
\newcommand{\h}{\hspace{0.30 cm}}

\newcommand{\n}{\nonumber}
\begin{document}
\title[Geometric approach to nonlinear coherent states using
the Higgs model for harmonic oscillator]
   {Geometric approach to nonlinear coherent states using
the Higgs model for harmonic oscillator}
\author{A. Mahdifar, R. Roknizadeh, and M. H. Naderi\\{Quantum Optics Group,
 University of Isfahan, Isfahan, Iran}}
\eads{ \mailto{a\_mahdifar@sci.ui.ac.ir},
\mailto{rokni@sci.ui.ac.ir}, \mailto{mhnaderi2001@yahoo.com}}
\begin{abstract}
In this paper, we investigate the relation between the curvature
of the physical space and the deformation function of the deformed
oscillator algebra using non-linear coherent states approach. For
this purpose, we study two-dimensional harmonic oscillators on
the flat surface and on a sphere by applying the Higgs modell.
With the use of their algebras, we show that the two-dimensional
oscillator algebra on a surface can be considered as a deformed
one-dimensional oscillator algebra where the effect of  the
curvature of the surface is appeared as a deformation function.
We also show that the curvature of the physical space plays the
role of deformation parameter. Then we construct the associated
coherent states on the flat surface and on a sphere and compare
their quantum
statistical properties, including quadrature squeezing and antibunching effect.\\

 \end{abstract}
  {\bf keywords}: coherent states on curved spaces, nonlinear coherent states,
  nonclassical properties\\

\section{Introduction}

   Coherent states (CSs) of the harmonic oscillator \cite{star} as well as generalized
   CSs associated with various algebras \cite{dstar,tav kla,tav ali(2000)}
   play an important role in various fields of physics, and in particular, they have
   found considerable applications in quantum optics. The former, defined as the
   eigenstates of the annihilation operator $\hat{a}$, have
   properties similar to those of the classical radiation field.
   The latter, on the contrary, may exhibit some nonclassical
   properties such as photon antibunching \cite{star2} or
   sub-Poissonian photon statistics \cite{dstar2} and squeezing
   \cite{cstar2}, which have given rise to an ever-increasing
   interest during the last decade.
   Among the generalized CSs, the so-called nonlinear CSs or f-deformed CSs \cite{star3}
   have attracted much attention in recent years, mostly because they exhibit
   nonclassical properties, such as amplitude squeezing and quantum interference
   \cite{dstar3,cstar3}. These states, which could be realized in the center-of-mass motion
   of an appropriately laser-driven trapped ion \cite{cstar3} and in a micromaser
   under intensity-dependent atom-field interaction \cite{chstar3}, are associated with
   nonlinear algebras and defined as the eigenstates of the annihilation operator of
   an f-deformed oscillator \cite{star3,tav manko}. Up to now, many quantum
   optical states such as q-deformed CSs \cite{tav manko}, the binomial states (or displaced excited CSs)
   \cite{tav rokni}, negative binomial states \cite{tav wang,tav wangfu} and photon-added
   photon-subtracted CSs \cite{tav siv(1999),tav naderi} and
   have been considered and studied as some examples of
   nonlinear CSs.\\
   One of the most important questions is the
   physical meaning of the deformation function in the nonlinear CSs.
   We have some theoretical and experimental data that consider
   the effect of curvature on the physical and optical properties
   of nano-structures (\cite{popov} and its references). On the other hand, it has been found
   \cite{naderi paian} that there is a close connection between the
   deformation function appeared in the nonlinear CSs algebraic
   structure and the non-commutative geometry of the configuration
   space. Furthermore, it is interesting to point out that our
   recent studies \cite{naderi nsr} have revealed strictly
   physical relationship between the nonlinearity concept
   resulting from f-deformation and some nonlinear optical
   effects, e.g., Kerr nonlinearity, in the context of atom-field
   interaction.
   These witnesses led us to this idea that there is a relation between deformation
   function and curvature of the physical space. In other words,
   we are intended to investigate that is it possible to find different deformation functions
   in the non-linear CSs corresponding to the geometry of space.\\
   In the present contribution, our main purpose is to find a geometrical
   interpretation for the f-deformation function based on the curvature parameter of
   physical space. For this purpose, we investigate the influence of
   the curvature of the space on the algebraic structure of the CSs and their
   physical properties within the framework of nonlinear coherent states approach.
   We first study the algebras of the
   two-dimensional harmonic oscillator on the flat space and on a non-flat space (sphere)
   in section two. In section three we show that these two oscillator algebras
   may be regarded as a one-dimensional f-deformed oscillator
   algebra. In particular, we find that the harmonic oscillator
   algebra on a sphere may be considered as a deformed version of
   oscillator algebra on the flat space. As an alternative point of view, in section
   four we show that the two-dimensional oscillator
   algebra on a sphere can be identified as a new type of deformed $su(2)$ algebra
   that in the limit of flat space it reduces to the standard $su(2)$
   algebra. In section five, we construct the corresponding CSs and examine their
   resolution of identity . Section six is devoted to the study of the quantum
   statistical properties of the constructed CSs. Especially, the
   influence of curvature of the physical space on their
   nonclassical properties is clarified. Finally, the summary and
   concluding remarks are given in section seven.

 \section{Deformed Algebra for two-dimensional harmonic oscillator}

  \subsection{Two-dimensional harmonic oscillator on flat space}

   The two-dimensional quantum harmonic oscillator
   in the flat space is described by the Hamiltonian
    \begin{equation}\label{bona 9}
    \hat{H}=\frac{1}{2}(\hat{p}_{x}^{2}+\hat{p}_{y}^{2})+\frac{1}{2}(\hat{x}^{2}
    +\hat{y}^{2})
    \end{equation}
   (in this paper we put $\hbar=m=\omega=1$). By using the Fradkin
   operators \cite{bona 54}
    \begin{equation}\label{bona 10}
    \hat{B}=\hat{S}_{xx}-\hat{S}_{yy}=(\hat{p}_{x}^{2}+\hat{x}^{2})
    -(\hat{p}_{y}^{2}+\hat{y}^{2}),\hspace{0.3 cm}
    \hat{S}_{xy}=\hat{p}_{x}\hat{p}_{y}+\hat{x}\hat{y},
    \end{equation}
   and angular momentum operator
    \begin{equation}\label{}
    \hat{L}=\hat{x}\hat{p}_{y}-\hat{y}\hat{p}_{x},
    \end{equation}
    one can define the following operators:
    \begin{eqnarray}\label{bona 14}
    &&
    \hat{n}=\frac{\hat{L}}{2}-u\hat{I},\n\\
    &&
    \hat{A}^{\dag}=\frac{1}{2}(\frac{\hat{B}}{2}+i\hat{S}_{xy}),\n\\
    &&
    \hat{A}=\frac{1}{2}(\frac{\hat{B}}{2}-i\hat{S}_{xy}),
    \end{eqnarray}
    where $u$ is a constant to be determined. From (\ref{bona 14}) it follows that
    $\hat{A}$, $\hat{A}^{\dag}$ and $\hat{n}$ satisfy the following closed algebraic
    relations \cite{bona}:
    \begin{eqnarray}\label{bona 15}
    &&
    [\hat{n},\hat{A^{\dag}}]=\hat{A}^{\dag}, \n\\
    &&
    [\hat{n},\hat{A}]=-\hat{A},\n\\
    &&
    [\hat{A},\hat{A}^{\dag}]=\Phi(\hat{H},\hat{n}+1)-\Phi(\hat{H},\hat{n}).
   \end{eqnarray}
   The structure function, $\Phi(E,n)=\frac{1}{4}[E^{2}-(2n+2u-1)^{2}]$, is a
   real definite positive function  for $n\geq 0$ and $\Phi(E,0)=0$. With this
   algebra, we can define the corresponding Fock space for each energy eigenvalue,
    \begin{eqnarray}\label{}
       &&
       \hat{H}|E,n\rangle=E|E,n\rangle,\n\\
       &&
       \hat{n}|E,n\rangle=n|E,n\rangle,\h n=0,1,2,\ldots,\n\\
       &&
       \hat{A}|E,0\rangle=0,\n\\
       &&
       |E,n\rangle=\frac{1}{\sqrt{[\Phi(E,n)]!}}(\hat{A^{\dag}})^{n}|E,0\rangle,
       \end{eqnarray}
      where, by definition
       \begin{equation}\label{}
       [0]!=1,\hspace{0.5 cm}
       [\Phi(E,n)]!=\Phi(E,n)[\Phi(E,n-1)]!\hspace{0.1 cm}.
       \end{equation}
      In the case of the discrete energy eigenvalues, for every eigenvalue $E$
      there is some degeneracy of dimension $N+1$ which is
      a root of the structure function. Therefore the dimensionality of
      the Fock space corresponding to that energy eigenvalue should be equal to
      $N+1$,
       \begin{eqnarray}\label{ali 1}
       &&
       \hat{H}|N,n\rangle=E_{N}|N,n\rangle,\h N=0,1,2,\cdots\n\\
       &&
       \hat{n}|N,n\rangle=n|N,n\rangle,\h n=0,1,2,\cdots,N.
       \end{eqnarray}
       This restriction, combined with
      the annihilation of the structure function for $n = 0$,
      determine the energy eigenvalues,
       \begin{equation}\label{bona 18}
       (E_{N})_{flat}\equiv(E_{N})_{f}=N+1,\h\h u=-\frac{N}{2},
       \end{equation}
      and therefore we find,
       \begin{equation}\label{bona 19.5}
       \Phi_{f}(E_{N},n)=n(N+1-n).
       \end{equation}

 \subsection{Two-dimensional harmonic oscillator on a sphere}

   There are several coordinate systems on a sphere which are useful
   generalisations of
   the Cartesian systems of Euclidean geometry. If we use the gnomonic projection,
   which is the projection onto the tangent plane from the centre of
   the sphere in the embedding space, and denote the Cartesian
   coordinates of this projection by $x_{i}$, the Hamiltonian
   of the harmonic oscillator on a sphere with curvature
   $\lambda=\frac{1}{R^{2}}$, may be written in the form \cite{higgs}
    \begin{equation}\label{}
    \hat{H}=\frac{1}{2}(\hat{\pi}^{2}+\lambda \hat{L}^{2})
    +\frac{1}{2}(\hat{x}^{2}+\hat{y}^{2}),
    \end{equation}
   where
    \begin{equation}\label{higgs 14a}
    \vec{\hat{\pi}}=\vec{\hat{p}}+\frac{\lambda}{2}\left[\vec{\hat{x}}(\vec{\hat{x}}.\vec{\hat{p}})
    +(\vec{\hat{p}}.\vec{\hat{x}})\vec{\hat{x}}\right],
    \end{equation}
   and
    \begin{equation}\label{higgs 5.5}
    \hat{L}^{2}=\frac{1}{2}\hat{L}_{ij}\hat{L}_{ij},\h \hat{L}_{ij}
    =\hat{x}_{i}\hat{p}_{j}-\hat{x}_{j}\hat{p}_{i}.
    \end{equation}
   Using the same method as
      for the flat harmonic oscillator, we obtain
       \begin{equation}\label{bona 25}
       (E_{N})_{sphere}\equiv(E_{N})_{s}=\sqrt{1+\frac{\lambda^{2}}{4}}(N+1)+\frac{\lambda}{2}(N+1)^{2},
       \h\h u=\frac{N}{2},
       \end{equation}
      and the structure function can be written as,
       \begin{equation}\label{bona 25.5}
       \Phi_{s}(E,n)=n(N+1-n)\left(\lambda(N+1-n)+\sqrt{1+\lambda^{2}/{4}}\right)
       \left(\lambda n+\sqrt{1+\lambda^{2}/{4}}\right).
       \end{equation}

 \section{Two-dimensional harmonic oscillator as a deformed harmonic oscillator}

      The annihilation and creation operators associated with an f-deformed
      harmonic oscillator algebra satisfy the following commutation relation\cite{tav
      manko}:
      \begin{equation}\label{tav 3}
       [\hat{A},\hat{A}^{\dag}]=(\hat{n}+1)f^{2}(\hat{n}+1)
       -\hat{n}f^{2}(\hat{n}),
       \end{equation}
      where
       \begin{eqnarray}\label{tav 1}
       &&
       \hat{A}=\hat{a}f(\hat{n})=f(\hat{n}+1)\hat{a},\n\\
       &&
       \hat{A}^{\dag}=f^{\dag}(\hat{n})\hat{a}^{\dag}=\hat{a}^{\dag}f^{\dag}(\hat{n}+1).
       \end{eqnarray}
      $\hat{a}$, $\hat{a}^{\dag}$ and $\hat{n} = \hat{a}^{\dag}\hat{a}$ are
      bosonic annihilation, creation and number operators,
      respectively and $f(\hat{n})$ is the deformation function.
      Ordinarily, the phase of $f$ is irrelevant and we can
      choose $f$ to be real and nonnegative, i.e. $ f^{\dag}(\hat{n}) =
      f(\hat{n})$.\\
      On the other hand, according to (\ref{bona 15}), we can write:
       \begin{equation}\label{ali 3}
       [\hat{A},\hat{A}^{\dag}]=\Phi(\hat{H},\hat{n}+1)-\Phi(\hat{H},\hat{n}).
       \end{equation}
      Thus if we work with a constant energy, $E=E_{N}$, then
      $\Phi(\hat{H},\hat{n})$ depends only on $\hat{n}$, and we can write
       \begin{equation}\label{ali 4}
       nf^{2}(n)=\Phi(E_{N},n).
       \end{equation}
      In the flat space where $\Phi_{f}(E_{N},n)$ is given by
      (\ref{bona 19.5}), we have
       \begin{equation}\label{ali 5}
      f_{f}(\hat{n})=\sqrt{(N+1-\hat{n})},
       \end{equation}
      while on the sphere where
      $\Phi_{s}(E_{N},n)$ is given by (\ref{bona 25.5})
      we have
       \begin{equation}\label{ali 6}
      f_{s}(n)=f_{f}(\hat{n})g(\lambda,n),
       \end{equation}
      where
       \begin{equation}
       g(\lambda,n)=\sqrt{\left(\lambda(N+1-n)
       +\sqrt{1+\lambda^{2}/{4}}\right)
       \left(\lambda n+\sqrt{1+\lambda^{2}/{4}}\right)}.
       \end{equation}
      It is obvious that in the flat limit, $g(\lambda,n)\rightarrow 1$
      and (\ref{ali 6}) reduces to (\ref{ali 5}).
      Therefore we can consider the two-dimensional harmonic
      oscillator algebra as a type of one-dimensional deformed
      harmonic oscillator with the deformation function $f(n)$ and also the sphere
      oscillator algebra as a deformed version of flat oscillator algebra with the deformation
      function $g(\lambda,n)$.\\

  \section{Two-dimensional harmonic oscillator algebra as a deformed
  $\frak{su}(2)$  algebra}
     In this section we show that the two-dimensional harmonic oscillator
     algebra can also be considered as a deformed $\frak{su}(2)$ algebra.
     For this purpose, we begin with the commutation relation
     \be
      [\hat{A}^{\dag},\hat{A}]=\hat{n}f^{2}(\hat{n})
       -(\hat{n}+1)f^{2}(\hat{n}+1),
     \ee
    [cf. eqation (\ref{tav 3})]. According to (\ref{ali 5}), the above
    relation in the flat space takes the following form:
     \be\label{a1}
      [\hat{A}^{\dag},\hat{A}]=2(\hat{n}-\frac{N}{2}).
     \ee
    If we now make the identifications
     \bea\label{id 1}
      (\hat{A}^{\dag})
      &\rightarrow&
      \hat{J_{+}}\n\\
      (\hat{A})
      &\rightarrow&
      \hat{J_{-}}\n\\
      (\hat{n}-\frac{N}{2})
      &\rightarrow&
      \hat{J_{0}},
     \eea
    we arrive at the standard $\frak{su}(2)$ algebra
    \begin{eqnarray}\label{a 2}
    &&
    [\hat{J_{0}},\hat{J_{\pm}}]=\hat{J_{\pm}}, \n\\
    &&
    [\hat{J_{+}},\hat{J_{-}}]=2\hat{J_{0}},
   \end{eqnarray}
   for which the eigen-states are $|j,m\rangle$s such that
   \be
   \hat{J_{0}}|j,m\rangle=m|j,m\rangle,\h -j\leq m \leq j.
   \ee
   If we now compare this relation with (\ref{ali 1}), we find that
   $n$s are positive while $m$s may take both negative and positive values.
   According to positivity of the values of $n$, which denote the number of
   excitation quanta, the identification of the above two algebras requires that
   a constant value $(\frac{2j}{2}=\frac{N}{2})$ is added to the eigenvalues $m$,
   so that they become positive. This is a reason for the identification
   $(\hat{n}-\frac{N}{2})\rightarrow\hat{J_{0}}$.\\
   On the other hand if we use the deformation function for the two-dimensional
   harmonic oscillator algebra on a sphere [eq.(\ref{ali 6})] and if we use the same
   identification as (\ref{id 1}), we have
    \begin{eqnarray}\label{a 2}
    &&
    [\hat{J_{0}},\hat{J_{\pm}}]=\hat{J_{\pm}}, \n\\
    &&
    [\hat{J_{+}},\hat{J_{-}}]=2\hat{J_{0}}\hspace{.1 cm} h(\lambda,N,\hat{J_{0}}),
    \end{eqnarray}
    where
    \be
    h(\lambda,N,\hat{J_{0}})=\left\{1+\lambda(1+\frac{\lambda}{4})^{1/2}(N+1)
    -\lambda^{2}\left[2\hat{J_{0}}^{2}-N(\frac{N}{2}+1)-\frac{1}{4}\right]\right\}.
    \ee
    It is clear that in the flat limit, $\lambda\rightarrow 0$,
    $h(\lambda,N,\hat{J_{0}})\rightarrow 1$ and the above deformed $\frak{su}(2)$
    algebra reduces to the standard non-deformed $\frak{su}(2)$ algebra.
    Thus, we conclude that two oscillator algebras (two W-H algebras) in
    the flat space with this approach can be regarded as an $\frak{su}(2)$
    algebra (like the schwinger model) and also these two
    algebras can be considered as a deformed
    $\frak{su}(2)$ algebra on a sphere. This deformed
    $\frak{su}(2)$algebra is one of the generalized deformed
    $\frak{su}(2)$ algebras defined in ref. \cite{bona su}.

  \section{Coherent states in the finite-dimensional Hilbert space}

      By using the definition of deformed creation and annihilation
      operators, [eqs.(\ref{tav 1})], and the deformation functions
      $f_{f}(\hat{n})$ and $f_{s}(\hat{n})$, we find,
       \begin{equation}\label{}
       \hat{A}|0\rangle=0=\hat{A}^{\dag}|N\rangle.
       \end{equation}
      Thus for each constant value of $N$
      (or constant value of energy $E_{N}$), we encounter with a finite dimensional Hilbert
      space. In this section, we are intended to construct coherent states
      associated with the flat space and sphere.\\
 \subsection{CSs on the flat space}
      On the flat space we have
       \begin{eqnarray}
       &&
       \hat{A}|n\rangle=\chi_{n}^{N}|n-1\rangle,\n\\
       &&
       \hat{A}^{\dag}|n\rangle=\chi_{n+1}^{N}|n+1\rangle,
       \end{eqnarray}
      where
       \begin{equation}
       \chi_{n}^{N}=\sqrt{n(N+1-n)}.
       \end{equation}
      We can make use of the formalism of constructing truncated
      coherent states \cite{mira 17} and define the finite-dimensional
      CSs as
       \begin{equation}\label{}
       |\mu\rangle_{f}=(1+|\mu|^{2})^{-N/2}\exp(\mu\hat{A}^{\dag})|0\rangle
       =(1+|\mu|^{2})^{-N/2}\sum_{n=0}^{N}\sqrt{\left(\begin{array}{c}
        N  \\
        n
        \end{array}\right)}\hspace{.1 cm}\mu^{n}|n\rangle,
       \end{equation}
      where $\mu$ is a complex number. It is seen that these
      CSs are analogous to the spin CSs that are constructed by using the $\frak{su}(2)$ algebra.

 \subsection{CSs on a sphere}

      From (\ref{tav 1}) and (\ref{ali 6}) we have
       \begin{eqnarray}
       &&
       \hat{A}|n\rangle=[g(\lambda,n)]\chi_{n}^{N}|n-1\rangle,\n\\
       &&
       \hat{A}^{\dag}|n\rangle=[g(\lambda,n+1)]\chi_{n+1}^{N}|n+1\rangle.
       \end{eqnarray}
      Therefore, as in the case of the flat space we can define the associated
      CSs on sphere,
       \begin{equation}\label{coh}
       |\mu\rangle_{s}=C\exp(\mu\hat{A}^{\dag})|0\rangle
       =C\sum_{n=0}^{N}\sqrt{\left(\begin{array}{c}
        N  \\
        n
        \end{array}\right)}[g(\lambda,n)]!\hspace{.1 cm}\mu^{n}|n\rangle,
       \end{equation}
      where
       \begin{equation}
       \frac{1}{C^{2}}=\sum_{n=0}^{N}\left(\begin{array}{c}
        N  \\
        n
        \end{array}\right)\{[g(\lambda,n)]!\}^{2}(|\mu|^{2})^{n}.
       \end{equation}
      It is found that for small values of $\lambda$, we have
       \begin{equation}
       g(\lambda,n)=1+\frac{\lambda}{2}(N+1)+o(\lambda^{2}),
       \end{equation}
      so that
       \begin{equation}
       [g(\lambda,n)]!=[1+\frac{\lambda}{2}(N+1)]^{n}.
       \end{equation}
      Thus in this limit we obtain
       \begin{equation}\label{limit}
       |\mu\rangle_{s}=C\sum_{n=0}^{N}\sqrt{\left(\begin{array}{c}
        N  \\
        n
        \end{array}\right)}[\mu(1+\frac{\lambda}{2}(N+1))]^{n}|n\rangle.
       \end{equation}
      In the other words, for obtaining
      the CSs, it is sufficient to make the replacement
      $\mu\rightarrow\mu(1+\frac{\lambda}{2}(N+1))$ for the CSs in the flat space in
      the large radius limit.

  \subsection{Resolution of identity}

      In this section we show that the CSs on the flat space and on
      a sphere form an overcomplete set. Since it is necessary to
      include a measure function $m(|\mu|^{2})$ in the integral, we
      require
       \be
       \int d^{2}\mu |\mu\rangle m(|\mu|^{2})\langle\mu|=\sum_{n=0}^{N}|n\rangle\langle
       n|=\mathbb{I}.
       \ee
      In the case of flat space,
       \bea
       \int d^{2}\mu |\mu\rangle_{f} m_{f}(|\mu|^{2}) _{f}\langle\mu|
       &=&
       \sum_{n=0}^{N}|n\rangle\langle n|\left(\begin{array}{c}
        N  \\
        n
        \end{array}\right)\int \frac{1}{2}d(|\mu|^{2})d\theta
        \frac{|\mu|^{2n}}{(1+|\mu|^{2})^{N}} m_{f}(|\mu|^{2})\n\\
        &=&
        \pi \sum_{n=0}^{N}|n\rangle\langle n|\left(\begin{array}{c}
        N  \\
        n
        \end{array}\right)\int_{0}^{\infty} d(|\mu|^{2})\frac{|\mu|^{2n}}
        {(1+|\mu|^{2})^{N}} m_{f}(|\mu|^{2})\n.
        \eea
       Thus we should have
        \be\label{int}
        \int_{0}^{\infty} d(|\mu|^{2})\frac{|\mu|^{2n}}
        {(1+|\mu|^{2})^{N}} m_{f}(|\mu|^{2})=\frac{1}{\pi \left(\begin{array}{c}
        N  \\
        n
        \end{array}\right)} .
        \ee
       The suitable choice for the measure function reads as \cite{radcliff}
        \be
        m_{f}(|\mu|^{2})=\frac{N+1}{\pi}\frac{1}{{(1+|\mu|^{2})^{2}}}.
        \ee
       In this manner the resolution of identity is
        \be
        \frac{N+1}{\pi}\int \frac{d^{2}\mu}{{(1+|\mu|^{2})^{2}}}
        |\mu\rangle_{f}\hspace{.1 cm} _{f}\langle\mu|=\mathbb{I}.
        \ee
       In order to examine the resolution of identity for the CSs on
       a sphere, we first introduce deformed Binomial expansion
       \be
       (1+x)_{\lambda}^{N}=\sum_{n=0}^{N}\left(\begin{array}{c}
        N  \\
        n
        \end{array}\right)_{\lambda}x^{n},
       \ee
       where by definition
       \be
       \left(\begin{array}{c}
        N  \\
        n
        \end{array}\right)_{\lambda}=\left(\begin{array}{c}
        N  \\
        n
        \end{array}\right)\{[g(\lambda,n)]!\}^{2}.
       \ee
       We see that when $\lambda\rightarrow 0$, $g(\lambda,n)\rightarrow
       1$ and the deformed Binomial expansion becomes the well-known Binomial
       expansion. With the use of this definition, we can write
       (\ref{coh}) as
       \be
       |\mu\rangle_{s}={(1+|\mu|^{2})_{\lambda}^{-N/2}}\sum_{n=0}^{N}\sqrt{\left(\begin{array}{c}
        N  \\
        n
        \end{array}\right)_{\lambda}}\mu^{n}|n\rangle.
        \end{equation}
       For the resolution of identity, we should have
        \be
        \int d^{2}\mu |\mu\rangle_{s} m_{s}(|\mu|^{2}) _{s}\langle\mu|
        =\sum_{n=0}^{N}|n\rangle\langle n|=\mathbb{I},
        \ee
        or
        \be
        \pi \sum_{n=0}^{N}|n\rangle\langle n|\left(\begin{array}{c}
        N  \\
        n
        \end{array}\right)_{\lambda}\int_{0}^{\infty} d(|\mu|^{2})\frac{|\mu|^{2n}}
        {(1+|\mu|^{2})^{N}_{\lambda}} m_{s}(|\mu|^{2})=\mathbb{I} .
        \ee
        If we define the corresponding measure as
         \be
         m_{s}(|\mu|^{2})=\frac{N+1}{\pi}\frac{1}{{(1+|\mu|^{2})_{\lambda}^{2}}},
         \ee
        and the deformed version of integral (\ref{int}) as
         \be
        \int_{0(\lambda)}^{\infty} d(|\mu|^{2})\frac{|\mu|^{2n}}
        {(1+|\mu|^{2})^{N}_{\lambda}} m_{s}(|\mu|^{2})=\frac{1}{\pi \left(\begin{array}{c}
        N  \\
        n
        \end{array}\right)_{\lambda}},
        \ee
        then we obtain the resolution of identity for the CSs on a sphere,
         \be
          \frac{N+1}{\pi}\int_{(\lambda)} \frac{d^{2}\mu}{{(1+|\mu|^{2})_{\lambda}^{2}}}
          |\mu\rangle_{s}\hspace{.1 cm} _{s}\langle\mu|=\mathbb{I}.
         \ee

\section{Quantum statistical properties of the flat and sphere CSs}
     Theoretically, we describe an ideal laser by the standard (Glauber) CSs.
     In particular, the photon-number distribution of this ideal
     laser, like the non-deformed CSs, is exactly Poissonian \cite{Naderi 38}.
     But the photon-number statistics of real lasers do not
     coincident with this description. It has been shown \cite{katriel}
     that, when the photon-number distribution is concerned, the q-deformed
     CSs are more suitable states for describing  nonideal
     lasers and other nonlinear interactions such as photons emitted by single-atom
     resonance fluorescence that we can say these photon states as
     deformed (or dressed) photon states.\\
     In the present section we shall proceed to study some quantum statistical
     properties of the constructed CSs , including mean number of photons, Mandel
     parameter and quadrature squeezing.

 \subsection{Photon-number distribution}

     The probability of finding $n$ quanta in the flat CSs,
     , i.e., its photon-number distribution is given by
       \begin{equation}
       P_{f}(n,\mu,N)=(1+|\mu|^{2})^{-N}\left(
       \begin{array}{c}
       N \\
       n
       \end{array}\right)
       \mu^{(2n)}.
       \end{equation}
     The mean number of photons in the state $|\mu\rangle_{f}$ is equal to
     \begin{equation}
      \langle\hat{n}\rangle_{f}=\langle\mu|\hat{a}^{\dag}\hat{a}|\mu\rangle_{f}
      =\sum_{n=0}^{N}nP_{f}(n,\mu,N).
     \end{equation}
     The photon-number distribution for the CSs on the sphere is given by
       \begin{equation}
       P_{s}(n,\mu,N,\lambda)=\frac{1}{\sum_{n=0}^{N}\left(
       \begin{array}{c}
       N \\
       n
       \end{array}\right)
       [g(\lambda,n)!]^{2}\mu^{(2n)}}\left(
       \begin{array}{c}
       N \\
       n
       \end{array}\right)
       [g(\lambda,n)!]^{2}\mu^{(2n)},
       \end{equation}
     and the mean number of photons in the state $|\mu\rangle_{s}$ is equal to
     \begin{equation}
      \langle\hat{n}\rangle_{s}=\langle\mu|\hat{a}^{\dag}\hat{a}|\mu\rangle_{s}
      =\sum_{n=0}^{N}nP_{s}(n,\mu,N).
     \end{equation}
     In Fig.1 we have plotted the variation of mean number of
     photons in the state $|\mu\rangle_{f}$ with respect to $\mu$ for
     different values of $N$. It is seen that for a given $N$,
     the mean number of photons is increased by increasing $\mu$
     and in the limit of $\mu\rightarrow\infty$ we have
       \begin{equation}
       {\lim_{\mu\rightarrow\infty}\langle\mu|\hat{n}|\mu\rangle_{f}\rightarrow}
       N,
       \end{equation}
     where $N$ is the dimension of the Hilbert space.\\
     Fig.2 displays the mean number of photons for the CSs on a
     sphere with respect to $\lambda$ for $N=10,20,30$ with $\mu=0.5$. As it is seen, for a fixed value of
     $N$, the mean number of photons is increased by increasing $\lambda$.
     It is in agreement with relation (\ref{limit}) for small $\lambda$.\\
     Since for the non-deformed coherent states the variance of the number operator
     is equal to its average, deviations from Poisson distribution can be measured
     with the Mandel parameter \cite{Naderi 37}
\begin{equation}
      M=\frac{(\Delta
      n)^{2}-\langle\hat{n}\rangle}{\langle\hat{n}\rangle},
      \end{equation}
     which is negative for a sub-Poissonian distribution (photon antibunching)
     and positive for a super-Poissonian distribution (photon-bunching).
     In Fig.3 we show the effect of curvature on the variation of
     Mandel parameter with $\mu=0.5$ and for different values of $N$.
     We see that when we go from flat space to sphere,
    photon-counting statistics of the CSs tends to sub-Poissonian more rapidly.
     In the other words, the CSs on the sphere show more nonclassical properties
     than the CSs on the flat space.\\

  \subsection{Quadrature squeezing}

     Let us consider the conventional quadrature operators $X_{1a}$ and $X_{2a}$
     defined in terms of nondeformed operators $\hat{a}$ and $\hat{a}^{\dag}$
     \cite{Naderi 38},
      \begin{equation}
      \hat{X}_{1a}=\frac{1}{2}(\hat{a}e^{i\varphi}+\hat{a^{\dag}}e^{-i\varphi}),\h
      \hat{X}_{2a}=\frac{1}{2i}(\hat{a}e^{i\varphi}-\hat{a^{\dag}}e^{-i\varphi}).
      \end{equation}
     The commutation relation for $\hat{a}$ and $\hat{a}^{\dag}$ leads to the following
     uncertainty relation,
       \begin{equation}
       (\Delta X_{1a})^{2}(\Delta X_{2a})^{2}\geq
       \frac{1}{16}|\langle[\hat{X}_{1a},\hat{X}_{2a}]\rangle|^{2}=\frac{1}{16}.
       \end{equation}
     In the vacuum state $|0\rangle$, we have $(\Delta X_{1a})_{0}^{2}
     =(\Delta X_{2a})_{0}^{2}=\frac{1}{4}$ and so
     $(\Delta X_{1a})_{0}^{2}(\Delta X_{2a})_{0}^{2}=\frac{1}{16}$. While it is
     impossible to lower the product $(\Delta X_{1a})^{2}(\Delta X_{2a})^{2}$
     below the vacuum uncertainty value, it is nevertheless possible to define squeezed
     states for which at most one quadrature variance lies below the vacuum value,
      \begin{equation}
      (\Delta X_{ia})^{2}<(\Delta X_{ia})_{0}^{2}=\frac{1}{4}\h\h(i=1\hspace{.1 cm}
      \mbox{or}\hspace{.1 cm} 2),
      \end{equation}
      or
      \begin{equation}
      S_{ia}(\varphi)=4(\Delta X_{ia})^{2}-1<0.
      \end{equation}
     Let us also consider here the deformed quadrature operators $\hat{X}_{1A}$
     a nd $\hat{X}_{2A}$ defined in terms of deformed operators $\hat{A}$ and
     $\hat{A}^{\dag}$,
      \begin{equation}
      \hat{X}_{1A}=\frac{1}{2}(\hat{A}e^{i\varphi}+\hat{A^{\dag}}e^{-i\varphi}),\h
      \hat{X}_{2A}=\frac{1}{2i}(\hat{A}e^{i\varphi}-\hat{A^{\dag}}e^{-i\varphi}).
      \end{equation}
     The commutation relation for $\hat{A}$ and $\hat{A}^{\dag}$ [eq. (\ref{tav 3}]
     leads to the following uncertainty relation:
       \begin{equation}
       (\Delta X_{1A})^{2}(\Delta X_{2A})^{2}\geq
       \frac{1}{16}|\langle[\hat{X}_{1A},\hat{X}_{2A}]\rangle|^{2}
       =\frac{1}{16}(\langle(\hat{n}+1)f^{2}(\hat{n}+1)-\hat{n}f^{2}(\hat{n})\rangle)^{2}.
       \end{equation}
       In this case, The condition of quadrature squeezing reads
       \begin{equation}
      (\Delta X_{iA})^{2}<\frac{1}{4}\langle(\hat{n}+1)f^{2}(\hat{n}+1)
      -\hat{n}f^{2}(\hat{n})\rangle\h\h(i=1 \hspace{.1 cm}\mbox{or}\hspace{.1 cm} 2),
      \end{equation}
      or equivalently,
      \begin{equation}
      S_{iA}=4(\Delta X_{iA})^{2}-\langle(\hat{n}+1)f^{2}(\hat{n}+1)\rangle
      +\langle\hat{n}f^{2}(\hat{n})\rangle<0.
      \end{equation}
     In Figures 4(a) and 4(b), respectively, we have plotted $S_{1a}$ and
     $S_{2a}$ with respect to $\varphi$ for N=10 and different values of $\lambda$.
     These figures clearly show that by increasing $\lambda$ the quadrature
     squeezing is enhanced.\\
     Figures 5(a) and 5(b), respectively, show $S_{1A}$ and $S_{2A}$ for N=10
     and different values of $\lambda$ . Here, we again find that by
     increasing $\lambda$ the degree of squeezing is enhanced.
     However, in comparison with the non-deformed quadratures, deformed
     quadratures exhibit stronger squeezing and increasing of $\lambda$
     leads to much pronounced squeezing. In summary, we can result that
     by going from the flat space to sphere the CSs
     exhibit increased nonclassical properties.

 \section{Summary and Concluding Remarks}

    In this paper, we have searched for a relation between the deformation
    function of the f-deformed oscillator algebra and two-dimensional harmonic
    oscillator on the flat space and on sphere. We have found that we could
    consider two-dimensional harmonic oscillator algebra as a deformed
    one-dimensional harmonic oscillator algebra. We have obtained two deformation function
    corresponding to the flat and sphere harmonic oscillators and
    we have also shown that sphere deformation function is a $g$-deformed with
    respect to the flat deformation function.
    Furthermore, we have found that the two-dimensional harmonic oscillator algebra
    on a sphere may be considered as a deformed $su(2)$ with a deformation
    function corresponding to the curvature of space that in the flat limit
    tends to unity. Then we have constructed the CSs for these spaces and studied their
    quantum statistical properties. The results show that the curvature of physical
    space leads to the enhancement of nonclassical properties of the
    states.\\
    We are now searching for a physical model that generates our coherent states
    on the flat and spherical surfaces. At the present step what we can
    conjecture is that photons on the sphere may be interpreted physically as
    dressed photons \cite{katriel} that are equivalent to bare photons
    plus a physical object (like a physical field).
    If we find that model, we can provide reasonable answer to the relevant questions about
    the physical meaning of the photons on the curved space. The work is in
    progress.\\
    Furthermore, the present contribution may be considered as the first
    step for finding the relation between nonlinearity of CSs
    and curvature of the physical space (here we have worked only with sphere).
    As another direction of the work, we have started on generalizing
    our approach to other non-flat surfaces, e.g. the surface with negative curvature,
    in order to find a complete relation between geometry and deformation function
    together with its physical implications.

\section*{Acknowledgement}
    The authors wish to thank The Office of Graduate Studies of The University of Isfahan
    for their support.

 \vspace{1cm}
 % ===============================================================================
 {\bf FIGURE CAPTIONS:}

  {\bf FIG. 1.} Mean number of photons in the state $|\mu\rangle_{f}$
  versus $\mu$, the dotted corresponds to $N=10$, the dashed to $N=20$
  and the solid curve to $N=30$.

  {\bf FIG. 2.} Mean number of photons in the state $|\mu\rangle_{s}$
  versus $\lambda$ for $\mu=0.5$, the dotted corresponds to $N=10$, the dashed to $N=20$
  and the solid curve to $N=30$.

  {\bf FIG. 3.} Mandel parameter versus $\lambda$ for $\mu=0.5$, the solid corresponds
   to $N=10$, the dashed to $N=20$ and the dotted curve to $N=30$.

  {\bf FIG. 4a.} $S_{1a}$ versus $\varphi$ for $N=10$ and $\mu=0.1$, the dotted corresponds
  to $\lambda=0.0$, the dashed to $\lambda=0.05 $ and the  solid curve to $\lambda=0.1 $.

  {\bf FIG. 4b.} $S_{2a}$ versus $\varphi$ for $N=10$ and $\mu=0.1$, the solid corresponds
  to $\lambda=0.0$, the dashed to $\lambda=0.05 $ and the  dotted curve to $\lambda=0.1 $.

  {\bf FIG. 5a.} $S_{1A}$ versus $\varphi$ for $N=10$ and $\mu=0.1$, the solid corresponds
  to $\lambda=0.0 $, the dashed to $\lambda=0.05 $ and the  dotted curve to $\lambda=0.1 $.

  {\bf FIG. 5b.} $S_{2A}$ versus $\varphi$ for $N=10$ and $\mu=0.1$, the solid corresponds
  to $\lambda=0.0 $, the dashed to $\lambda=0.05 $ and the  dotted curve to $\lambda=0.1 $.


\section*{References}
\begin{thebibliography}{100}
  \bibitem{star} R. J. Glauber, Phys. Rev. {\bf 136}, 2529 (1963);
                 R. J. Glauber, Phys. Rev. {\bf 131}, 2766 (1963);
                 R. J. Glauber, Phys. Rev. Lett. {\bf 10}, 84 (1963).

  \bibitem{dstar} A. P. Perelomov, Generalized Coherent States and
                  Their Applications (Springer, Berlin, 1986).

  \bibitem{tav kla} J. R. Klauder and B. S. Skagerstam, Coherent States, Applications
                    in Physics and Mathematical Physics, (Singapoore, World Scientific, 1985).

  \bibitem{tav ali(2000)} S. Twareqe Ali, J-P. Antoine, and J-P. Gazeau, Coherent
                          States, Wavelets and Their Generalizations,
                          (Springer-Verlag, New York, 2000).

\bibitem{star2} H. J. Kimble, M. Dagenais and L. Mandel, Phys.
                  Rev. Lett. {\bf 39}, 691 (1977).

\bibitem{dstar2} M. C. Teich and B. E. A. Saleh, J. Opt. Soc. Am.
                 B {\bf 2}, 275 (1985).

\bibitem{cstar2} C. K. Hong and L. Mandel, Phys. Rev. Lett.
                 {\bf 54}, 323 (1985);
                 L. A. Wu, H. J. Kimble, J. L. Hall and H Wu,
                 Phys. Rev. Lett. {\bf 57}, 2520 (1986).

\bibitem{star3} A. I. Solomon, Phys. Lett. A {\bf 196}, 29 (1994);
                J. Katriel and A. I. Solomon, Phys. Rev. A
                {\bf 49}, 5149 (1994);
                P. Shanta, S. Chaturvdi, V. Srinivasan and R. Jagannathan,
                J. Phys. A: Math. Gen. {\bf 27}, 6433
                (1994).

\bibitem{dstar3} R. L. de Matos Filho and W. Vogel, Phys. Rev. A
                 {\bf 52}, 4214 (1995).

\bibitem{cstar3} W. Vogel and R. L. de Matos Filho, Phys. Rev. A {\bf 54}
                 4560 (1996).

\bibitem{chstar3} M. H. Naderi, M. Soltanolkotabi and R.
                  Roknizadeh, Eur. Phys. J. D. {\bf 32}, 397 (2005).

\bibitem{tav manko} V. I. Man'ko, G. Marmo,E. C. G. Sudarshan and F. Zaccaria,
                    Physica Scripta {\bf{55}}, 528 (1997).

\bibitem{tav rokni} R. Roknizadeh and M. K. Tavassoly, J. Phys. A: Math. Gen.
                       {\bf 37}, 5649 (2004).

\bibitem{tav wang} J. Liao, X. Wang, L-A Wu and S-H Pan, J. Opt. B:
                   Quantum Semiclass. Opt. {\bf 2}, 541 (2000).


\bibitem{tav wangfu} X-G Wang and H-C Fu, Commun. Theor. Phys.
                     {\bf 35}, 729 (2001).

\bibitem{tav siv(1999)} S. Sivakumar, J. Phys. A: Math. Gen. {\bf 32}, 3441 (1999).

\bibitem{tav naderi} M. H. Naderi, M. Soltanolkotabi and R. Roknizadeh,
                    J. Phys. A: Math. Gen. {\bf 37}, 3225 (2004).

\bibitem{popov} V. N. Popov, New J. Phys. {\bf 6}, 17 (2004).

\bibitem{naderi paian} M. H. Naderi, Ph.d. thesis, university of
                       Isfahan, Iran (2004) (unpublished)

\bibitem{naderi nsr} M. H. Naderi, M. Soltanolkotabi and R.
                     Roknizadeh, J. Phys. Soc. Japan, {\bf 73},
                     2413, (2004)

\bibitem{bona 54} D. M. Fradkin, Am. J. Phys. \underline{33}, 207 (1965).

\bibitem{bona} D. Bonatsos, C. Daskaloyannis, K. Kokkotas, Phys.
                 Rev. A {\bf 50}, 3700 (1994).

\bibitem{higgs} P. W. Higgs, J. Phys. A: Math. Gen. {\bf 12}, 309 (1979).

\bibitem{Naderi 37} L. Mandel and E. Wolf. Optical coherence and quantum optics
                      (Cambridge University Press, Cambridge 1995).

\bibitem{Naderi 38} M.O. Scully and M.S. Zubairy. Quantum optics (Cambridge
                    University Press, Cambridge 1997).

\bibitem{radcliff} J. M. Radcliffe, J. Phys. A: Math. Gen. {\bf 4}, 313 (1971).

\bibitem{mira 17} L. M. Kuang, F. B. Wang, and Y. G. Zhou, Phys.
                    Lett. A {\bf 183}, 1 (1993).

\bibitem{bona su} D. Bonatsos, C. Daskaloyannis and P. Kolokotronis,
                    J. Phys. A: Math. Gen. {\bf 26}, 871 (1993).

\bibitem{katriel} J. Katriel, A. I. Solomon, Phys. Rev. A {\bf 49}, 5149
                   (1994).

\end{thebibliography}
\end{document}